# *Enhancement of Hydroxyapatite Dissolution through Krypton Ion Irradiation*

Hui Zhu [1], Dagang Guo [1,*], Hang Zang [2,*], Dorian A. H. Hanaor [3], Sen Yu [1,4], Franziska Schmidt [3], Kewei Xu [1]

[1] *State Key Laboratory for Mechanical Behavior of Materials, School of Material Science and Engineering, Xi'an Jiaotong University, Xi'an 710049, PR China*

[2] *School of Nuclear Science and Technology, Xi'an Jiaotong University, Xi'an 710049, PR China*

[3] *Fachgebiet Keramische Werkstoffe/Chair of Advanced Ceramic Materials, Institut für Werkstoffwissenschaften und -technologien, Technische Universität Berlin, Hardenbergstraße 40, 10623 Berlin, Germany*

[4] *Shaanxi Key Laboratory of biomedical metal materials, Northwest Institute for Non-ferrous Metal Research, Xi'an 710016, China*

*Corresponding author. Prof. DaGang Guo. Tel.: +86 29 82668614;    guodagang@mail.xjtu.edu.cn

*Corresponding author. A. Prof. Hang Zang.; Tel.: +86 29 82665915;    zanghang@mail.xjtu.edu.cn

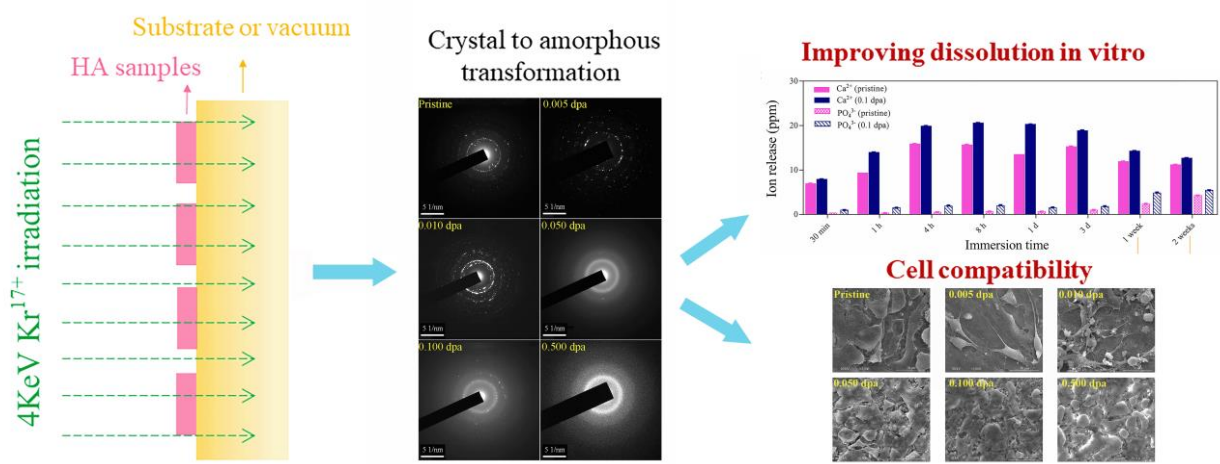

## Abstract

Hydroxyapatite (HA) synthesized by a wet chemical route was subjected to heavy ion irradiation, using 4 MeV Krypton ions ($Kr^{17+}$) with ion fluence ranging from $1\times10^{13}$ to $1\times10^{15}$ ions/cm$^2$. Glancing incidence X-ray diffraction (GIXRD) results confirmed the phase purity of irradiated HA with a moderate contraction in lattice parameters, and further indicated irradiation-induced structural disorder, evident by a broadening of diffraction peaks. High-resolution transmission electron microscopy (HRTEM) observations indicated that the applied Kr irradiation induced significant damage in the hydroxyapatite lattice. Specifically, cavities were observed with their diameter and density varying with irradiation fluences while a radiation-induced crystalline-to-amorphous transition with increasing ion dose was identified. Raman and X-ray photoelectron spectroscopy (XPS) analysis further indicated the presence of irradiation-induced defects. Compositional analysis of pristine and irradiated materials following immersion in Tris (pH 7.4, 37℃) buffer showed that dissolution *in vitro* was enhanced by irradiation, reaching a peak for 0.1dpa. We examined the effects of irradiation on the early stages of the mouse osteoblast-like cells (MC3T3-E) response. A cell counting kit-8 assay (CCK-8 test) was carried out to investigate the cytotoxicity of samples, and viable cells can be observed on the irradiated materials.

# 1. Introduction

Hydroxyapatite (HA), $Ca_{10}(PO_4)_6(OH)_2$, exhibits natural biocompatibility and bioactivity as it is the dominant inorganic component of human bones. As proposed by Hench LL et al. [1], both bioactivity and biodegradability are the core requirements of third-generation biomedical materials. However, the extremely slow degradation rate of HA cannot match the growth rate of newly formed bone *in vivo* [2]. It has been reported that a bone-bonding region at HA interfaces is achieved through extensive degradation/recrystallization events, which lead to better bone-ingrowth fixation [3, 4]. That is, degradation of calcium phosphate-based biomaterials has a strong impact on their osteoconductive properties [5, 6]. Consequently, improving the degradation rate of this material has emerged as a key objective in recent biomaterial research. As reported, the *in vivo* degradation of Ca–P biomaterials is mainly determined by its dissolution performance [7, 8]. Various approaches have been taken towards enhancing the dissolution performance of HA , such as increasing the material's porosity [9, 10], reducing the Ca/P ratio [11], decreasing the grain size or crystallinity [12] and ion substitution [13]. A.E. Porter et al. demonstrated through *in vitro/vivo* experiments that the dissolution of silicon (Si)-substituted HA starts and propagates around structural defects [13-15]. Similarly , the dissolution of natural bone apatite such as human enamel is also driven by crystal defects, including abundant point defects, dislocations, stacking faults, and an amorphous grain boundary phase [16, 17]. In general, it has been estalished that the amount and nature of defects in biological apatite governs its reactivity and solubility [16, 18]. Methods applied to date to enhance apatite dissolution, through defect engineering or otherwise, have met with limited success, motivating the examination of further materials design approaches.

Heavy ion irradiation is a promising and effective technique to engineer the morphology and crystallinity of nanostructured materials and nanocomposites and thus modify their physicochemical properties. Through elastic and inelastic collisions with the nuclei and electrons of target atoms, energetic ions induce the displacement of atoms from their lattice positions resulting in defects and structural transformations [19]. B.D Begg et al. reported a "pyrochlore-to-fluorite" transformation in conjunction with amorphization triggered by irradiation [20]. This treatment was found to lead to a significant change in the material's dissolution rate.

Only a few studies have examined the influence of ion irradiation on hydroxyapatite and its performance as a biomaterial. Reports have shown that ion irradiation of HA may increase its surface roughness, wettability, mechanical properties, cell adherence and drug loading/release characteristics [21-26]. Improved levels of in-vitro bioactivity in HA have been imparted while avoiding cytotoxicity by means of swift heavy ion irradiation by 100 MeV oxygen ions [26], and 125 MeV silicon ions [25]. These results have indicated that high energy ion beam irradiation may serve as an effective method to enhance the biocompatibility and bioactivity of HA. Relative to more commonly used techniques such as doping or substitution, ion irradiation has the advantage of altering the material's crystal structure without altering its stoichiometry. However, in general, the mechanisms through which ion irradiation affects the performance of bioceramics remain unclear and the dissolution performance of irradiated HA is yet to be examined. Gleaning further insights into the role of ion bombardment in biomaterials engineering necessitates direct examination of irradiation induced defects.

In the current study, we combine a visual method—High Resolution Transmission Electron Microscopy (HRTEM), with X-Ray Diffraction (XRD) and spectroscopic analysis to study radiation-induced structural modification and the crystalline-to-amorphous transition of HA by 4 MeV $Kr^{+17}$ ions. The effects of irradiation on the *in vitro* dissolution performance and *in vitro* cell responses of HA are further examined. Through these studies, we provide the basis for a novel crystal structure engineering strategy to improve the degradation of HA.

## 2. Experimental

### 2.1. Materials

Stoichiometric HA [$Ca_{10}(PO_4)_6(OH)_2$] was synthesized by a wet chemical method, previously described in greater detail [27]. Briefly, the pH of 0.5 M $Ca(NO_3)_2·4H_2O$ and $(NH_4)_2HPO_4$ (SINOPHARM Group Co.Ltd, China) water solution were adjusted to 10.0 and 9.0 respectively by ammonia. The diammonium phosphate solution was then added slowly under stirring to the solution of calcium nitrate to yield a molar ratio of Ca/P =1.67, and the solution was sealed following pH adjustment to 10.5. The reaction mixture was vigorously shaken at 50 °C for 48 h and then aged for 24 h. Precipitates were collected after washing. The paste was dried at 100 ˚C under vacuum and was ground manually. After sintering in a vacuum oven at 900 °C for 2 h, the obtained material was ground to yield a fine 200-mesh powder suitable for TEM analysis.

### 2.2. Sample preparation

Samples were prepared in different forms prior to irradiation treatment. For TEM analysis, nanoparticles of HA that had been ultrasonically dispersed in solvent were deposited on copper TEM grids. For X-ray diffraction (XRD), Raman and XPS analysis, HA coatings were fabricated as follows: HA powder in ethanol was sonicated and stirred overnight to form a well dispersed colloid. Fused silica substrates were pre-treated using a dilute HF solution and rinsed in DI water in turn. The substrates were cleaned in acetone, ethanol and water ultrasonically, and then dried at 60 °C. HA dispersions were deposited on the sheets followed by spinning at 8000 RPM for 40 s. After spinning, the thin coatings were dried for 10 min on a hot plate at 60 °C. Thickness measurements, using Confocal Laser Scanning Microscopy (CLSM), confirmed a uniform thickness of approximately 2 μm (Fig. 1a) from this process. For dissolution and cell studies, HA discs (2 mm height and 13 mm diameter) were uniaxially compacted from pure powders. All samples were fixed on a copper holder for perpendicular penetration of $Kr^{17+}$ ions (Fig. 1b).

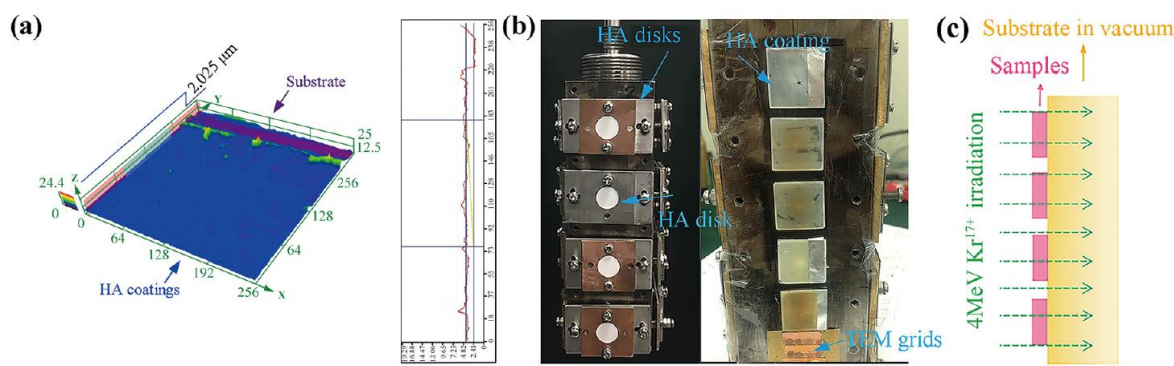

**Fig. 1**. (a) 3D surface morphology gradient map of a typical HA coating on a fused silica substrate. (b) Sample fixing setup for ion irradiation. (c) A diagram of irradiation process.

## 2.3. Ion irradiation

Irradiation experiments were carried out at the Institute of Modern Physics, Chinese Academy of Science (IMPCAS). Using the 320 kV Multi-discipline Research Platform for Highly Charged Ions Materials, HA samples were irradiated by 4 MeV $Kr^{17+}$ ions at room temperature, with fluences of $1\times10^{13}$, $2\times10^{13}$, $1\times10^{14}$, $2\times10^{14}$ and $1\times10^{15}$ ions/cm$^2$ as shown in Fig.1(c). The energy and fluences used here were selected to impart significant alteration of local structure with no residual radioactivity [28]. The local ionization and displacement densities per ion as a function of depth in HA under these irradiation conditions were determined from full cascade Monte Carlo calculations using the SRIM-2008 (Stopping and Range of Ions in Matter) code, with output shown in Fig. 2. Default displacement energies of 25, 25, 28 and 10 eV for Ca, P, O and H, respectively, were applied here [29, 30]. The relative amount of damage arisen from nuclear collisions is given as displacements-per-atom (dpa: the average times each target atom has been displaced) [31]. Fig. 2 reveals that Kr ion irradiation at $1.0\times 10^{15}$ ions/cm$^2$ leads to an irradiation damage of 0.5 dpa on the surface and a peak irradiation induced damage of 1.7 dpa accompanied by a peak Kr deposition concentration of about 0.02 %. As indicated by the pink regime on the left side of Fig. 2, irradiation penetration depth range is much larger than the TEM foil thickness (100 nm), and hence defects are expected to be uniformly introduced along the specimen thickness. Krypton implantation effects can be neglected within this region, For the 2 μm coatings on silica substrates, the thickness of which is indicated by the yellow regime in Fig. 2, the damage varies with depth, however, this does not impede the analyses presented here. The final samples were denoted in Table 1.

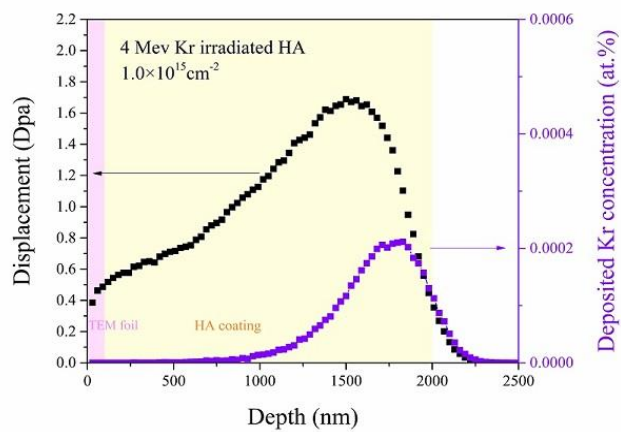

**Fig. 2**. SRIM simulation of the irradiation induced damage variations and Kr ion distribution with depth. Irradiation on the Cu foil and bulk Cu sample. Pink area in the figure represents the 100-nm TEM foils and yellow areas represent the 2 μm HA coatings.

**Table 1** Correspondence of dpa to ion fluences

| Sample label | Displacements-per-atom | Ion fluence ions/cm$^2$ |
|---|---|---|
| S0 | 0 | 0 |
| S1 | 0.005 dpa | $1\times10^{13}$ |
| S2 | 0.01 dpa | $2\times10^{13}$ |
| S3 | 0.05 dpa | $1\times10^{14}$ |
| S4 | 0.1 dpa | $2\times10^{14}$ |
| S5 | 0.5 dpa | $1\times10^{15}$ |

## 2.4. Material Characterization

Phase compositions of HA thin films before and after irradiation were determined by Glancing incidence X-ray diffraction (GIXRD) with CuKα radiation, using a scan speed of 2 °/min and increment step of 0.077. On the basis of absorption calculations [32], a fixed incident angle of 1.5° was selected in order to examine a targeted penetration depth of 2 μm, corresponding to the thickness of the film. To image irradiation induced defects, samples, before and after irradiation treatment, were examined in a JEM-2100F High Resolution Transmission Electron Microscope (HRTEM) with an acceleration voltage of 200 kV, TEM imaging is known to induce lattice damage in HA materials. For this reason, imaging was conducted as briefly as possible to minimize electron irradiation influence. Energy-dispersive X-ray spectroscopy (EDS) and Selected Area Electron Diffraction (SAED) were performed on the irradiated samples as well. To examine bonding, Raman spectroscopy was employed over the range of shifts between 250 and 4,000 cm$^{-1}$ in a backscattering geometry with 325 nm laser source on a confocal Horiba Jobin-Yvon HR800 spectrometer. The X-ray photoelectron spectroscopy (XPS) analyses were achieved by a Al Kα radiation for the excitation.

## 2.5. In vitro dissolution

The dissolution characteristics of pristine and irradiated materials were determined by a static immersion test. Each disc shaped sample was coated on one side with impervious lacquer to yield a constant exposed area. Then, discs were fixed by wax to the tip of a Teflon tube and immersed in tris (hydroxymethyl)-aminomethane) (VWR. Co, US) solution with a pH of 7.4. The samples were sealed in PP Teflon containers at 37 °C for up to 14 days. The specimen area (mm$^2$) to solution volume ratio (ml) (SA/V) was 10:1 following the standard GB/T16886.14-2003[33]. The solute concentrations of $Ca^{2+}$ and $PO_4^{3-}$ were measured from aliquots taken at various intervals by inductively-coupled-plasma (ICP) emission spectroscopy [IRIS Advantage (Thermo Scientific Ltd)]. The cumulative ion release up to 14 days were calculated and plotted subsequently. For TEM characterization of dissolution effects, powders scraped from the irradiated HA coatings were immersed in deionized water

## 2.6. In vitro cell culture studies
### 2.6.1. Cell culture

The biological performance of the as-prepared and irradiated HA discs were evaluated by using mouse osteoblast-like MC3T3 cells, obtained from cell banks of the Chinese Academy of Sciences (Shanghai, China). The culture medium was composed of Alpha Modified Eagle's medium (α-MEM, Hyclone, USA), 10 % (v/v) fetal bovine serum (FBS, Zhejiang Tianhang Biotechnology Co.,Ltd, China) and 1 % penicillin-streptomycin (Sigma-Aldrich, USA).

### 2.6.2. Cell Attachment and Proliferation Assay

All disks were sterilized by gamma rays (Co-60) of 25 kGy for 24 h. Approximately 80 mL of cell suspension (50000 cells/cm$^2$·mL$^{-1}$ for cell attachment and 25000 cells/cm$^2$·mL$^{-1}$ for cell proliferation, respectively) was poured on top of the discs gently. After 1 h at 37 °C, the remaining medium was carefully added to the wells to cover the surface (1.0 mL). The medium was replaced every second day.

Cell attachment and morphology was observed after incubating for 24 h by scanning electron microscopy (SEM, FEI Quanta 400, Netherlands). Samples were washed twice with PBS solution, and then fixed for 30 min in 0.1 M sodium-phosphate-buffered solution with 2.5 % glutaraldehyde (Sinopharm, Shanghai, China). After washing twice with PBS solution, samples were dehydrated in ethanol with ascending concentrations (30, 50, 70, 90, 95, 100 % (v/v)) for 10 min each, followed by immersion in pure hexamethyldisilazane for 10 min. Then they were dried in a desiccator overnight. The dried specimens were coated with gold for SEM.

Cell proliferation was evaluated after

seeding cells followed by incubation for 1, 3, and 7 days following a cell counting kit-8 (CCK-8, Dojindo, Japan) assay. Briefly, the sample disks were first washed with PBS solution gently and then were carefully removed to a new 24-well plate. Afterwards, each well with sample were flooded with 360 μL cell culture medium and 36 μL CCK-8 solution. After incubation at 37 °C for 3 h, 100 μL solution from each well of the 24-well plate was transferred to another 96-well cell culture plate quickly. The optical density (OD) was measured by a spectrophotometric microplate reader (GENios, Germany) with 450 nm wavelength. The OD values at days 3 and 7 were normalized to those at day 1.

### 2.6.3. Statistical analysis

One-way Anova was adopted for statistical analysis. Quintuple samples were used in each experiment to obtain the standard deviation. Statistically significant differences were considered at $p<0.05$.

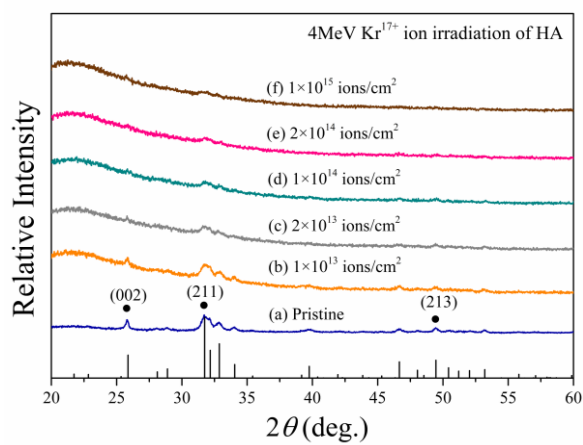

**Fig. 3**. GIXRD spectra of the irradiated and unirradiated HA coatings. Ion fluence: (a) $1\times10^{13}$ to (f) $1\times10^{15}$ ions/cm$^2$.

## 3. Results

### 3.1. XRD study

GIXRD was applied to analyze the crystalline structure in thin films samples on fused silica substrates. The depth of damage in irradiated HA was estimated by SRIM as 2.5 μm (Fig. 2), meaning that the entire volume of the supported film experiences irradiation induced structural modification. Fig. 3 shows the diffraction peaks for HA thin films irradiated with 4 MeV Kr for various ion fluences. Compared with the JCPDS standard cards (#72-1243) of pure HA phase, no other phases were detected in irradiated materials. Notably, with increasing ion fluence, the intensities of (002) (211) and (213) diffraction peaks decreased and their widths increased, indicating lattice distortion. Furthermore, slight shifts of peaks to lower 2θ angles were observed. No impurity phases appeared after irradiation.

### 3.2. Formation of nanocavities during irradiation

To investigate local structure alteration, we directly irradiated HA nanoparticles deposited on TEM copper grids. Incident Kr ions passed completely through the grids. Fig. 4a shows HRTEM images of HA nanoparticles before and after irradiation. Unirradiated particles exhibited grains without any vacancies or defects. Irradiation induced cavities formed and developed with increasing dose. For S1 and S2 with low irradiation doses, significant cavities (mean diameter 5.6 nm to 6.7 nm) formed (marked by arrows). For an irradiation fluence of $1\times10^{14}$ ions/cm$^2$ (S3), the original cavity diameter increased further to 8.2 nm approximately. Additionally, a second class of smaller cavities (approximately 2.4 nm) were also observed, particularly between the surface and the main layer. Continued Kr ion irradiation led to a doubling of the diameter of both the larger (14.1 nm) and smaller (6.9 nm) cavities, reaching a peak at a fluence of $2\times10^{14}$ ions/cm$^2$ (S4). However, in specimens irradiated further to $1\times10^{15}$ ions/cm$^2$ (S5), the mean diameters of the cavities were found to be significantly smaller. The evolution of cavity size with Kr fluence is shown in Fig. 4b. Interestingly, the surface density of cavities exhibited a similar trend, showing a maximum in S5. Following the highest ion fluence of $1\times10^{15}$ ions/cm$^2$, small cavities were found to be primarily located in close proximity to free surfaces , as such free surfaces may serve as a sink for point defects [34]. The EDS taken from S3 in Fig. 4d shows no Kr peaks, and a weakened oxygen peak compared to the unirradiated HA atomic arrangement of $Ca_{10}(PO_4)_6(OH)_2$ (Fig. 4c).

### 3.3. Transmission electron microscopy analysis of ion irradiation induced "crystal-amorphous" transition

A series of bright field images of HA nanoparticles following irradiation at increasing fluences are shown in Fig. 5. The TEM photographs of Kr–irradiated HA nano particles indicate that Kr ion irradiation resulted in significant nanostructural modification. Fig. 5a corresponds to a typical TEM image of pristine material (unirradiated) showing a pure single crystal without evidence of defects, as evident from the undistorted lattice fringes. Small mottled contrast appears in S1 following an irradiation dose of $1\times10^{13}$ ions/cm$^2$, as shown in Fig. 5b. Some lattice fringes are discontinuous and they often manifest as noticeable edge dislocations, which are marked in Fig. 5b and its inverse fast Fourier transform (IFFT) in Fig. 5c. The lattice spacing from Fig. 5c is found to be 0.281 nm, which matches closely the d-spacing of (211) planes of the HA lattice. Most regions are dominated by regular lattice stacking sequences, which reflects small disordering embedded in an otherwise undistorted crystal. Results indicate that at this irradiation dosage the apatite structure is largely intact, in agreement with XRD data. Following a dose of $2\times10^{13}$ ions/cm$^2$ in S2, lattice stacking was still present in most regions; besides, greater distortion and twisting of the lattice fringes was evident. Planar defects such as grain boundaries constituted the dominant features (Fig. 5d). Additionally, in S2 shown in in Fig. 5e, amorphous bands were observed interrupting the crystal, which could originate from screw dislocations and be coordinated in a single parallel network [35]. After $1\times10^{14}$ ions/cm$^2$ in S3, the degree of crystallinity continued to decrease and the mottled regions formed a continuous amorphous network (Fig. 5f). The original crystal is broken into isolated tiny crystallites and the lattice fringes were disordered around the amorphous regions. Fig. 5g indicates that amorphization proceeds inwards under continued irradiation to $2\times10^{14}$ ions/cm$^2$ in S4. A few crystalline islands were found, embedded in the amorphous matrix. Fast Fourier Transformation (FFT) patterns, shown in the inset of Fig. 5g, show that amorphization is indeed not complete. Finally, when the fluence exceeded $1\times10^{15}$ ions/cm$^2$ or 0.5 dpa, the lattice image indicates a nearly fully amorphous state as shown in Fig. 5h. Only mottled contrasts are observed without any lattice stacking sequences. Furthermore, the SAED patterns displayed in Fig. S-1 confirmed the "crystal-to-amorphous" transformation triggered by irradiation as well. Overall, TEM images indicate a clear progressive increase in damage in HA nanoparticles by Kr irradiation: pristine samples exhibit regular structures, while with increasing ion irradiation the level of damage gradually increases until complete amorphization.

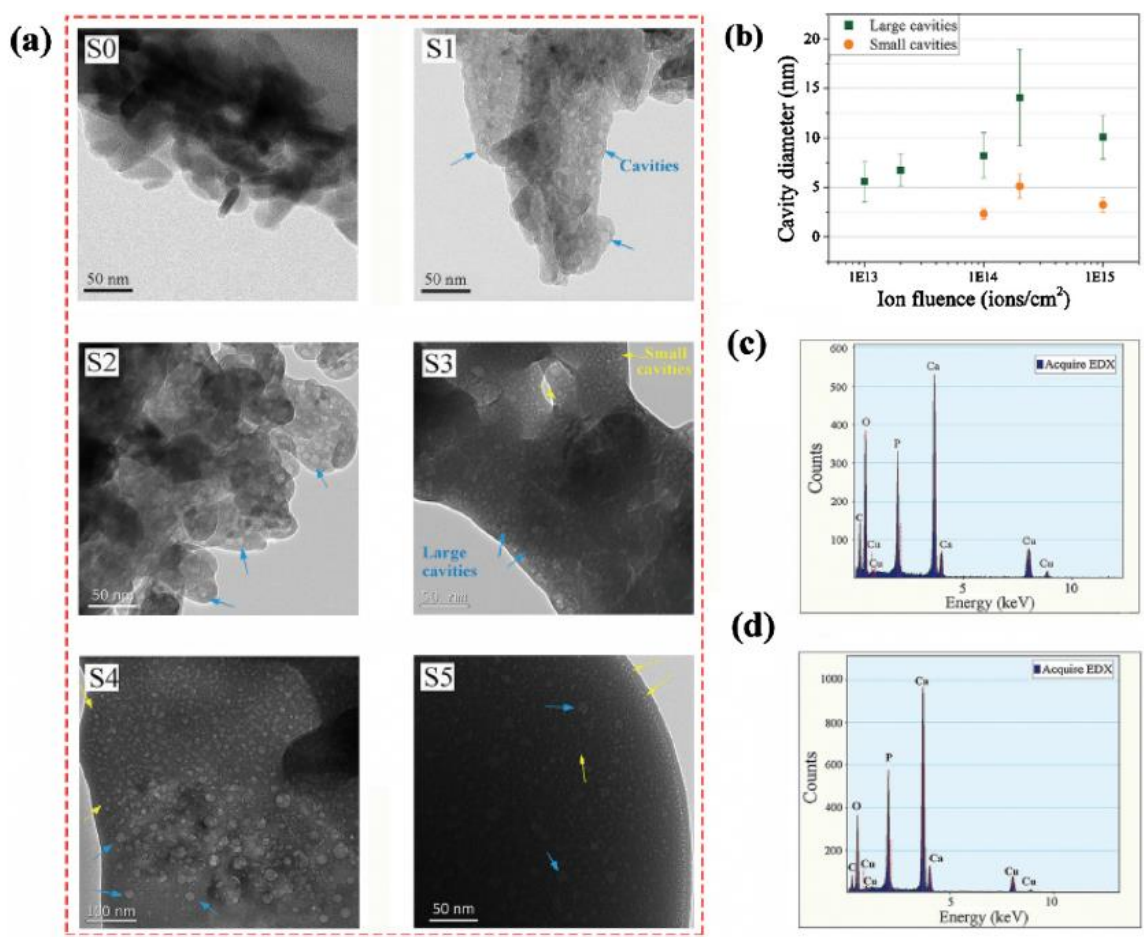

**Fig. 4**. (a) Cavities formation and evolution in the HA nanoparticles after irradiation with 4 MeV Kr$^{17+}$ ions with increasing irradiation fluence at room temperature. (b) Cavity size variations upon the heavy ion fluence from TEM observations. (c) EDS spectrum taken from S0. (d) EDS spectrum taken from S3.

*3.4. Raman spectroscopy*

Raman spectra for apatite have been extensively studied. According to those studies, Raman active vibrations can be attributed to the following phonon modes: ~962 cm$^{-1}$ to the $\nu_1$ symmetric stretching vibration of phosphate anions (the most intense mode in the apatite spectrum), ~430 and ~453 cm$^{-1}$ to the $\nu_2$ symmetric out-of-plane bending modes, ~1,052 and ~1,081 cm$^{-1}$ to the $\nu_3$ antisymmetric stretching modes, and ~591 and ~608 cm$^{-1}$ to $\nu_4$ anti-symmetric bending, and the bond stretching mode associated with the OH group is detected as a sharp peak at 3572 cm$^{-1}$ [36]. Fig. 6a shows the overall HA Raman spectra before and after irradiation. Raman shifts for different spectral regions are presented in Fig. 6b-e. Decreased peak intensities and broadened full widths at half maximum (FWHM) are evident with increasing ion fluence. Notably, the $\nu_3$ phonon peak and OH stretching almost disappear at $1\times10^{14}$ ions/cm$^2$ and the previously separated and sharp modes disappear. Moreover, a red-shift could be observed with rising ion fluence as shown in Fig. 6c. Those observations are all indicative of amorphization. The strongest $\nu_1$ band (962 cm$^{-1}$) (inset in Fig. 6c) was further analyzed by deducing the evolution of its peak intensity, position and FWHM with deconvolution and least squares fitting of classical Poisson law [37, 38]. Fig. 6f illustrates this peak's intensity trend evolving with ion fluence i. Intensities decayed drastically from 2770.54 for the pristine material to 391.45 following a fluence of $1\times10^{15}$ ions/cm$^2$. A Poisson's relation was employed here to fit the damage fraction in irradiated hydroxyapatite:

$$F_d = B[1 - \exp(-A\Phi t)] \quad (1)$$

Where $A=\pi R^2$ presents the cross section for amorphization and B represents the maximal possible amorphization for hydroxyapatite [39, 40]. According to the fitting results, the radius (R) can be calculated as 1.15±0.27, and the value of B can be obtained as 0.90±0.02. Based on this calculation, even the highest fluence cannot lead to a complete amorphization of hydroxyapatite, which is in disagreement with the above TEM results. However, it should be noted that it is very difficult to establish a straightforward correlation between damage in individual nanoparticles and the 'integrated' Raman spectrum due to the limited penetration depth and different Raman scattering cross-sections [31]. Fig. 6g-h show the evolution of peak positions and FWHMs of the ν1 band with flthe e. Those trends are similar to the changes of peak position described above. Compared to the non-irradiated sample, a maximum peak shift of 2.3 cm$^{-1}$ was observed at the fluence of $1\times10^{15}$ ions/cm$^2$. The shift could be attributed to the presence of residual stresses in the material induced by the growing number of cavities and defects in the irradiated samples. [37].

It should be noted that a new broad band, centered around ~945 cm$^{-1}$ can be observed after Kr$^{17+}$ ion irradiation, becoming pronounced above the fluence of $2\times10^{13}$ ions/cm$^2$ (Fig. 6c). It grows with the increment of ion fluence. This broad band ($\nu_{1b}$) is formed from the ion irradiated amorphous cores [41]. The area ratio of $\nu_{1b}$ to the $\nu_1$ symmetric stretching vibration of phosphate anions at 962 cm$^{-1}$ was widely used for quantifying the amount of amorphization in different samples [38]. Following the same method, spectra were fractionally fitted by applying the Lorenz equation with the result reported in Fig. 6e. As anticipated, the area ratio of ν1b to ν1 band increased gradually with ion fluence. For instance, it increased from 0.25±0.04 to 6.06±0.33 from $1\times10^{13}$ to $1\times10^{15}$ ions/cm$^2$, indicating a growing trend of amorphization.

*3.5. XPS analysis*

O 1s X-ray photoelectron spectroscopy of the samples and its fitting results before and after irradiation were also examined in Fig. 7 to support the inferences drawn from Raman spectroscopic studies. The first peak at 530.84 eV (I) is caused by O-P bonding and the second at 532.16 eV (II) is related to oxygen bonded to OH$^-$ groups [42], or oxygen vacancies [43]. After ion irradiation, the binding energy of lattice oxygen (O-P) shifted slightly from 530.84eV to 530.38eV, and its area decreased by 29.79 %. Whereas, the non-lattice oxygen (OH) binding energy shifted from 532.16 eV to 531.60, with its area increasing by 173.71% (see Supplementary Information; Table S1).

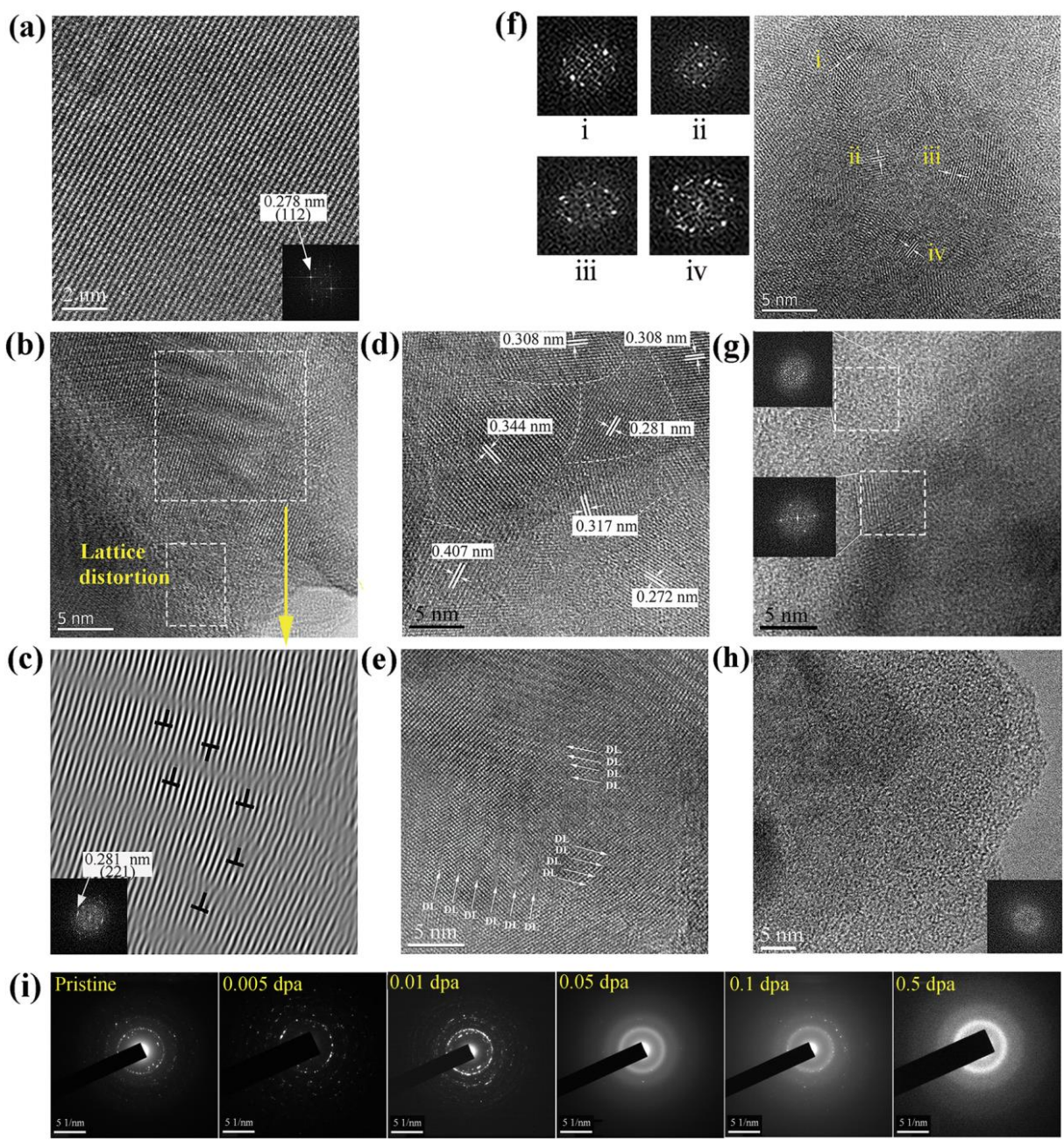

**Fig. 5**. Typical TEM images of the HA nano particles before and after Kr irradiation. (a) unirradiated, (b) 1×10$^{13}$ ions/cm$^2$ (0.005 dpa) and (c) its inverse fast fourier transform, (d) and (e) 2×10$^{13}$ ions/cm$^2$ (0.01 dpa), (f) 1×10$^{14}$ ions/cm$^2$ (0.05 dpa), (g) 2×10$^{14}$ ions/cm$^2$ (0.1 dpa), (h) 1×10$^{15}$ ions/cm$^2$ (0.5 dpa) (i) SAED patterns of the sampelsirradiated with various fluences.

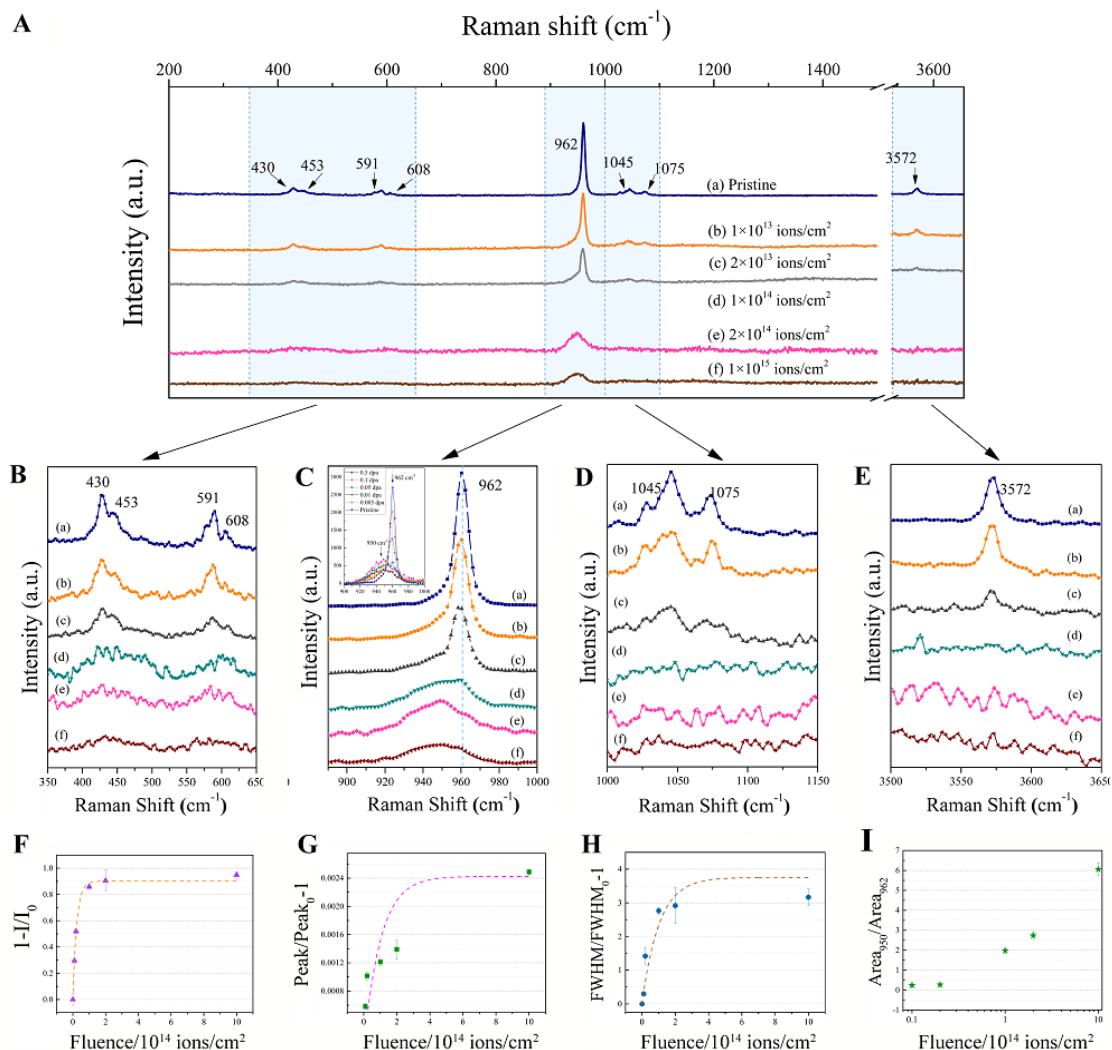

**Fig. 6**. Raman analysis: Spectra of samples irradiated at different doses are shown (a) over the full range of shifts, and in the shift ranges of (b) 125 to 3650 cm$^{-1}$, (c) 890 to 1000 cm$^{-1}$, (d) 1000 to 1150 cm$^{-1}$, (e) 3500 to 3650 cm$^{-1}$. Decay of this band's intensity, change of its position, and the increase of its breadth as indicated by FWHMs are plotted as a function of ion fluence in (f) (g) and (h) respectively. (i) Shows the area ratio of the radiation-induced broad $v_{1b}$ band (950 cm$^{-1}$) to the $v_1$ band (962 cm$^{-1}$).

*3.6. Dissolution study*

Fig. 8 shows the change in calcium and phosphate release in the Tris buffered solutions over 14 days of immersion. The curves in Fig. 8a-b show the impact of immersion time and Kr$^{17+}$ ion irradiation fluence on Ca$^{2+}$ and PO$_4^{3-}$ release. All curves present an initial fast release of Ca$^{2+}$ and PO$_4^{3-}$ over the first 4 hours immersion. The subsequent variation of Ca and P concentration is as follows: for pristine HA, Ca$^{2+}$ concentration decreased after immersion for 4 hours, and then fluctuated slightly until 3 days' immersion before decreasing. For samples subjected to lower irradiation doses (S1 and S2), Ca concentrations decreased after 8 hours soaking in Tris buffer and then remained stable from 1 to 3 days. Subsequently, significant decrease occurred. However for the samples under higher irradiation doses (S3-5), Ca concentrations remained constant from 8 to 24 hours immersion, and then dropped intensely. All samples showed similar trends of their $PO_4^{3-}$ release, which grows over the whole immersion in Tris buffer solution.

In addition, Fig. 8a-b shows that the ion release of irradiated HA was higher than pristine HA during their immersion in Tris buffer solution. Both of the $Ca^{2+}$ and $PO_4^{3-}$ concentration increased with the growing ion dose, and then reached maximum in S4 at 0.1 dpa ($2\times10^{14}$ ions/cm$^2$) irradiation. The $Ca^{2+}$ and $PO_4^{3-}$ concentration of S4 increased by 51.30 % and 133.57 % respectively compared to the pristine one for 1 day's immersion (Fig. 8c). However, when the ion dose increased to 0.5 dpa, the ion release dropped.

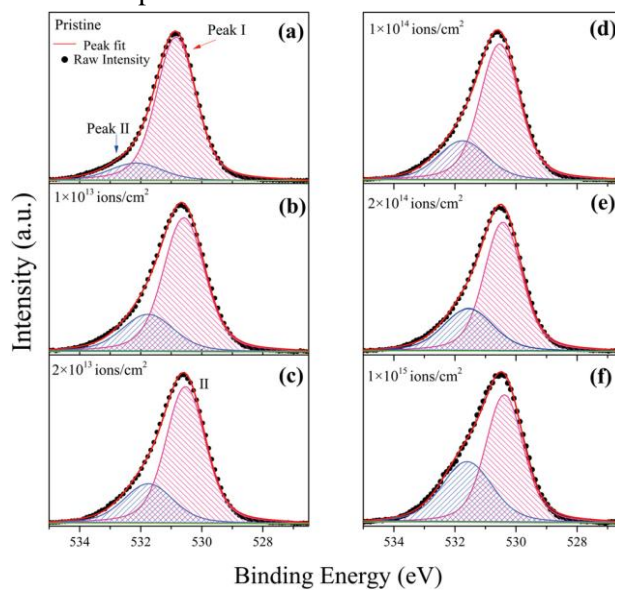

**Fig. 7**. XPS O 1s spectra for samples with increasing irradiation dose. (a) unirradiated, (b) $1\times10^{13}$ ions/cm$^2$ (0.005 dpa), (c) $2\times10^{13}$ ions/cm$^2$ (0.01 dpa), (d) $1\times10^{14}$ ions/cm$^2$ (0.05 dpa), (E) $2\times10^{14}$ ions/cm$^2$ (0.1 dpa), (f) $1\times10^{15}$ ions/cm$^2$ (0.5 dpa). In each subfigure, black markers show the data points, the solid red line shows the fitted profile and the two shaded areas show the deconvoluted contributions of the two binding energies.

To further understand the nanostructural variations of unirradiated and irradiated HA during dissolution, HRTEM observations were carried out for S0 and S4 respectively (Fig. 8d). After being immersed in deionized water for 1 week, unirradiated HA presented a mottled grain surface which indicate the dissolution has started from this region, while there was no sign of dissolution for the internal grain. In contrast, dissolution in irradiated HA initiated preferentially from defect structures inside grains, including grain boundaries and dislocations, with a corrugated disordered appearance. After 2 weeks, the dissolution continued around those mottled regions on grain surfaces in unirradiated HA, showing a lamellar-like peeling dissolution mode. For the irradiated counterpart, dissolution developed around the grain distortions and grain surfaces as well, suggesting a deeper level of dissolution.

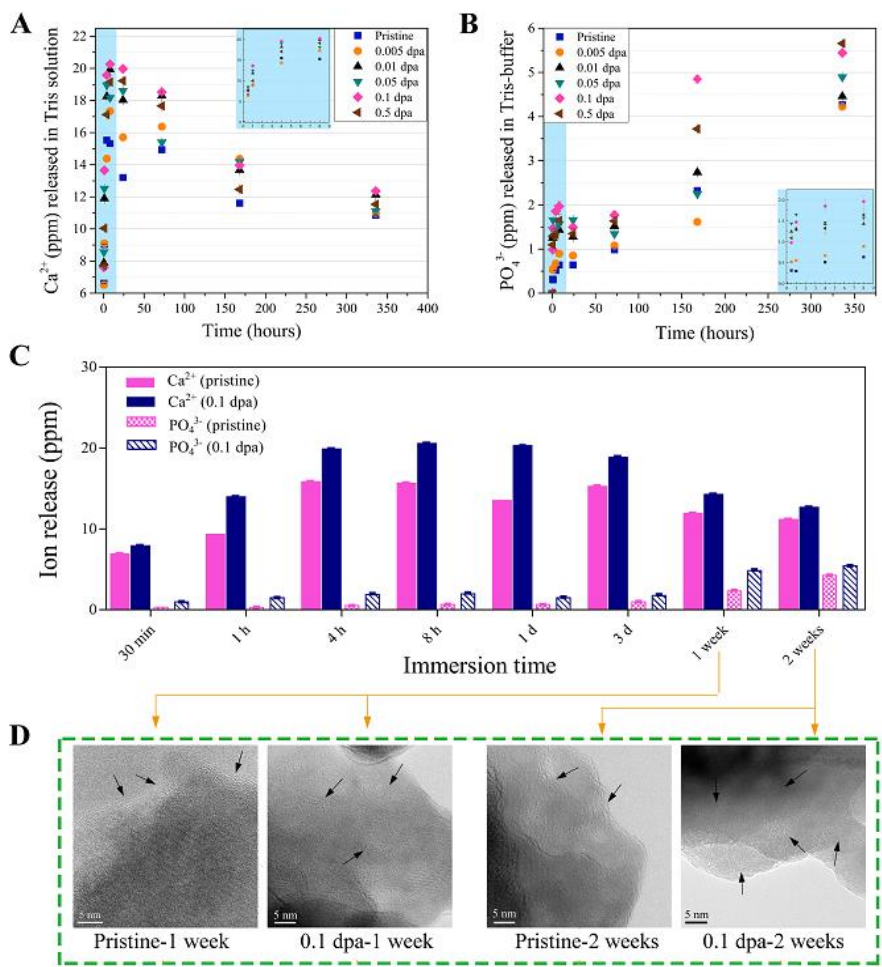

**Fig. 8**. (a) Calcium and (b) phosphate ions concentrations released into Tris-buffer solution from HA disks irradiated by 4 MeV $Kr^{17+}$ ions with different fluence, with respect to immersion duration. Insets: enlarged view from 30 min to 8 h immersion. (c) Comparison of the ion release of pristine sample and 0.1 dpa irradiated samples. (d) HRTEM pictures of the pristine and 0.1 dpa irradiated samples after immersion for 1 and 2 weeks. Arrows indicate the dissolution starting points.

*3.7. Cell culture study*

The morphology of attached cells on sample surfaces was observed by SEM to assess the cytocompatibility of un- /irradiated samples at 24 h after cell seeding (Fig. 9. A-f). MC3T3 cells attached and spread on all sample surfaces. After 24 hours of culture, dense layers of cells with some filopodia extension were observed covering the entire unirradiated HA surface. For S1, Cells were connected to each other with the longest filopodia extensions compared to other groups. Simultaneously, some apatite like grains can be detected on its surface. On S2 surface, a luxuriant growth and spread of cells can be seen with numerous filopodia, suggesting an good cell-materials attachment (Fig. 9c). Large numbers of spherical apatite particles were also observed as a result of mineralization. On the S3 and S4 irradiated surface, cells with unsatisfactory shapes could still interact with each other with shorter filopodia extension (Fig. 9d-e). A large quantity of apatite graduals formed beneath the cell layers. On the S5 surface (Fig. 9f), cells spread on the top of each other with a large density and intensive micro extensions.

Fig. 9g-h present the *in vitro* cytocompatibility results. Fig. 9g illustrates the numbers of attached cells on as-prepared and irradiated HA surface with increasing irradiation fluence after 24 hours' incubation. The OD values are widely used as a reliable measurement for the numbers of adherent cells on the specimen surface. In general, pristine HA had better cell attachment than irradiated HA (Fig. 9g, p< 0.05 compared to unirradiated sample). S3 and S4 have the minimal amount of adherent cells, indicating an unsatisfactory attachment. Fig. 9h presents the proliferation of MC3T3-E1 cells cultured on the as-prepared and irradiated HA surface as determined by CCK-8 assay. The irradiated sample with highest Kr ion fluence (0.5 dpa) exhibited a similar relative proliferation rate compared to the unirradiated counterpart at day 3 and day 7. However, the other irradiated groups had lower proliferation rate of MC3T3-E1 cells than unirradiated group, but no significant difference existed between them (p<0.05).

## 4. Discussion

This current work for the first time explores the utility of dissolution enhancement by irradiation towards the design of functional HA biomaterials and explores the mechanisms of this enhancement to provide a theoretical and experimental basis for its implementation.

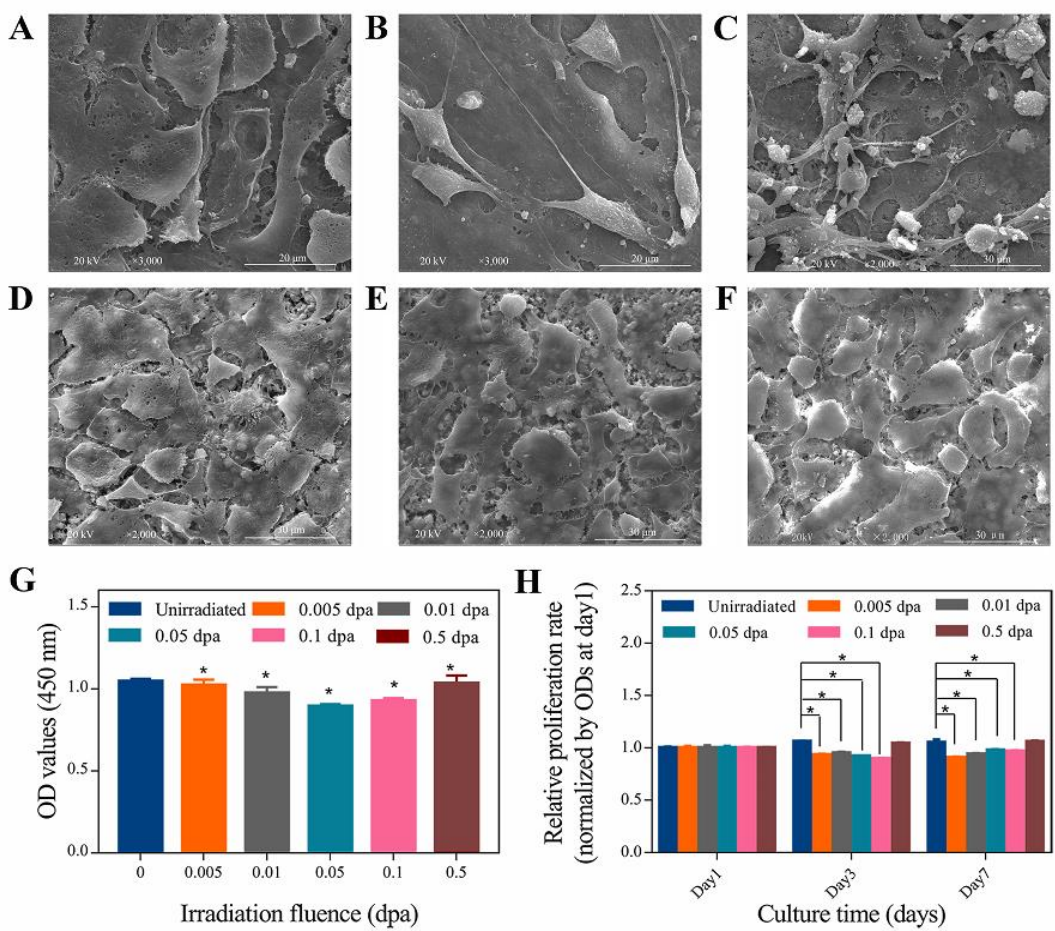

**Fig. 9**. Selected SEM images of cultured specimens ((a)-(f) $1\times10^{13}$ to $1\times10^{15}$ ions/cm$^2$) at 24 h after MC3T3-E1 were seeded on their surfaces; (g) cell adhesion of MC3T3-E1 cells on the specimens at 24 hours; (h) cell proliferation after cell seeding for 1, 3 and 7 days. The modified OD values at 3 and 7 days were normalized to those at 1 day. * denote p< 0.05.

During the passage of heavy ions through the material, energy dissipation occurs through two mechanisms : elastic collisions with the nuclei of target atoms, and inelastic collisions with electrons, both of which generate structural damage. The atoms in the target migrate and hit the surrounding atoms leading to collision cascades [31, 44]. The relationship of local structural damage and ion irradiation energy was obtained here by SRIM calculations, showing the utility of the applied treatment for modifying crystalline hydroxyapatite in micron-scale depth regimes. As a result of collisions, vacancies are formed due to the displacement of atoms from their original lattice sites. As vacancies form and coalesce, voids can be formed [45]. If insoluble atoms (He, Xe, Kr, etc.) aggregate within irradiation-produced vacancies, small bubbles will be formed [46]. The cavities observed in this study (see Fig. 4) should be categorized as voids, rather than bubbles as the SRIM calculations shown in Fig. 2 indicate that following the highest irradiation level, negligible Kr implantation is expected within the geometry of the TEM foil (100 nm thickness) and EDS spectra show no evidence of Kr peaks (see Fig. 4 of 0.05 dpa irradiated HA). This suggests that the voids observed by HRTEM arise from the agglomeration of irradiation-induced vacancies, formed through atomic displacement cascades, rather than pressurized Kr bubbles. Supporting this, the deconvolution of XPS O1s spectra shows a decreasing ratio of lattice oxygen to non-lattice oxygen contributions (Fig.. 7). This indicates the formation of irradiation induced oxygen vacancies, which may be the origin of the observed voids [47]. Furthermore, the shift of the lattice component of O1s to lower binding energies confirms that defect formation is associated with electronic charge transfer，and the formation of oxygen vacancies[48]. The formation of oxygen vacancies is further supported by a notable decrease in the oxygen signal observed in EDS relative to Ca and P, although these data are not conducive to quantitative interpretation.

The absence of significant void formation in pristine materials here suggests that beam damage did not play a role in present observations. In addition, the variation trend of mean void diameters with increasing ion fluences in Fig. 5 might be explained by the total swelling and the effect of surface energy [49].

Diffraction patterns from GIXRD (Fig. 2) clearly show an increasing amorphization with higher ion irradiation dose. The large baseline hump and the absence of significant peaks following the final dose indicate near complete disruption of long range order in the HA crystal structure. No secondary phase formation is evident. Comparison with a related study involving 125 MeV $Si^{9+}$ ion irradiated HA suggests 4 MeV $Kr^{17+}$ irradiation yields a similar outcome [25]. The shift to lower $2\theta$ angles implies lattice expansion accompanies the introduction of irradiation induced defects and structure disruption as observed elsewhere [32].

The observations of the present work lead to insights into the progression of structural defects in hydroxyapatite. It is postulated that oxygen vacancies are introduced by elastic collisions in the lattice. The bonding energy of oxygen ions in the vicinity of vacancies is thus diminished and they are more readily vaporized, leading to further vacancies [50]. With increasing fluence, a network dislocation structure forms, which has been reported to act as an unsaturable sink for irradiation induced point defects [51]. The presence of unfaulted dislocations and network dislocations, as seen here, is an indicator of significant point defect mobility (Fig. 5b). As irradiation progresses, oxygen deficiency increases and damage is enlarged [50].

The results of the present and previously published experiments indicate that introducing heavy ions can promote the formation and expansion of grain boundaries (Fig. 5d-e), which could provide the space for the movement of oxygen vacancies [52]. This is also considered as the beginning stage of polygonization. The relative disorder increases steadily with irradiation due to the accumulation of defect clusters and saturates after a crystal to amorphous transition. The threshold dose for complete amorphization of HA at room temperature under 4 MeV $Kr^{17+}$ ion irradiation obtained in the present study was around 0.5 dpa, which is consistent with earlier research on apatite and other ceramic materials [53]. It should be noted that HRTEM observation has its limits considering the very localized and thin regions of sample, nevertheless, diffraction results confirm the observed amorphous transition seen in these images. A summary of irradiation damage progression is presented in Table 2.

Raman spectra have been widely used to verify structural variations in irradiated apatite [37, 38, 41, 54], and similar results were obtained in this study. In general, irradiation induced defects and short-range disorder lead to a loss of intensity and broaden of bands [38] (Fig. 6). The blue shift of peak positions could further be due the residual stress caused by irradiation [55]. Notably, the growing are ratio of ν1b to ν1 at 964 $cm^{-1}$ facilitates the quantitative measurement of lattice damage, which is difficult to achieve by XRD [37]. Nevertheless, a total amorphization were not detected even with the highest fluence from Raman results. A possible reason could be that Raman spectra are determined by local bonding structure rather than lattice spacing, this result may indicate the preservation of a consistent short range order retained following irradiation induced disruption of the crystalline lattice.

The significant crystal structural variations in HA, arising from 4 MeV $Kr^{17+}$ ion irradiation, were found to enhance the material's *in vitro* dissolution throughout the immersion process. Enhanced solubility is a natural consequence of the higher surface and bulk free energy induced by defects and amorphization. As can be seen in Fig. 8a, maximum $Ca^{2+}$ release took place in the first 24 h for all samples, before subsequently declining, which might be a result of the consumption of $Ca^{2+}$ ions in the growth of apatite layers on sample surfaces[56]. Generally, it is seen here that the dissolution of HA, indicated by the release of $Ca^{2+}$ and $PO_4^{3-}$ was enhanced with an increasing density of crystal defects such as dislocations, grain boundaries and the expanding amorphous domains triggered by irradiation. This is in consistent with studies into bone apatite, that have shown crystal defect sites are initiation points for dissolution processes *in vivo/ vitro* [15, 57]. The ion concentration in bulk solution increased with irradiation damage in the range 0.005~0.1 dpa, and decreased with further damage to 0.5 dpa. Interestingly, this follows the same trend of the diameter variation of voids produced by irradiation, shown in Fig. 4. This result demonstrates the significance of nanoscale porosity in enhancing dissolution and further suggests that ion vacancies may act as critical dissolution nuclei, as proposed in earlier reports [58].

To evaluate biocompatibility, in this study, cells were directly seeded on HA disks. Our results demonstrated that $Kr^{17+}$ ion irradiation has slight negative effect on the cell adhesion and cell proliferation on HA samples, however, cell attachment of S1, S2, S5 at 24 h and cell proliferation of S3, S4, S5 at 7 d are comparable with those of S1, which demonstrated that they have similar effects on cell activities. Considering the reported cell viability on HA synthesized by aqueous precipitation method, the cytocompatibility of the irradiated HA samples in this work were acceptable as well [59]. Contact angles were measured on all the disk samples for cell study with no significant differences detected (Fig. 10). Therefore, we postulate that the enhanced response of S5 are attributable to the different surface chemistry and topography of materials, as well as higher levels of amorphization [60]. While the somewhat impaired responses of S1-S4 can be understood in light of reports that elevated pH levels arising from higher dissolution rates introduce a cytotoxic effect for cell proliferation and attachment [24].

Interestingly, the variation of attached cell numbers with irradiation fluence showed a similar trend to that of void diameters. A previous study has suggested that the presence of nanoscale voids in alumina exposes more cell-adhesive epitopes, promoting enhanced cell attachment [61]. Thus, it may be the case that the response of cellular adhesion proteins to the nanotopology of irradiation modified HA is the reason for the observed cell attachment differences in our study.

It should be noted that in most cases, ion irradiation only influences surface material to a depth of several microns, without changing the bulk material's structure. Nevertheless, this strategy still has a broad significance towards biomedical applications, including coatings and composite systems, where interface modification has an important role to play. Overall results here demonstrate that heavy ion irradiation is applicable for tailoring both dissolution and biocompatibility, motivating further investigations into this complex relationship.

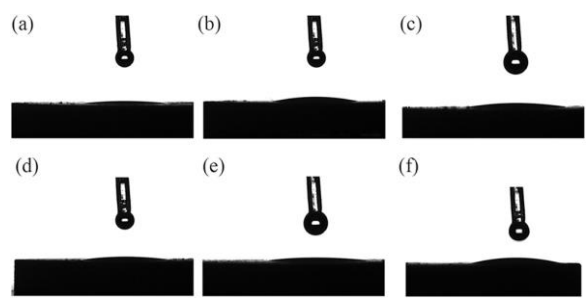

**Fig. 10**. Contact angle measurement of pristine and irradiated specimens with increasing irradiation fluences ((a)-(f) $1\times10^{13}$ to $1\times10^{15}$ ions/cm$^2$).

**Table 2.** A summary of irradiation damage progression of HA nanoparticles after irradiation.

| Irradiation dose | HA crystal stuructures |
|---|---|
| 0 | Perfect lattice arrangement |
| 0.005 dpa | Few dislocations and small mottled contrast |
| 0.01 dpa | Grain boundaries and numerous dislocation lines. Considerable number of disorders |
| 0.05 dpa | The original single crystal is broken into isolated crystallites with continuous amorphous phase. Lattice fringe are still visable. |
| 0.1 dpa | Few crystalline regions embedded in the amorphous matrix |
| 0.5 dpa | Completely amorphous HA particles. |

## 5. Conclusions

Through a combination of analytical approaches, heavy ion irradiation is shown here to be an effective technique for crystal structure engineering in hydroxyapatite. Irradiation induces structural disorder, leading to the loss of crystallinity, formation of defects and nucleation of cavities with dose dependent size and density. With sufficient heavy ion fluence full amorphisation of hydroxyapatite can be achieved, accompanied by lattice oxygen deficiencies and nanoscale voids. The effects of ion irradiation on the atomic structure and microstructure of hydroxyapatite enhance the material's performance in terms of biodegradation as measured by in-vitro dissolution, with a maximal effect observed following an irradiation dose calculated to yield 0.1 displacements per atom. This treatment would not impart considerable cytotoxicity as measured by MC3T3 cell attachment and proliferation on samples surfaces. Results here shed further light on the mechanisms through which ion irradiation can facilitate the engineering of hard tissue repairing materials.

**Acknowledgement**

This work was supported by the Science Challenge Project [No: TZ2018004], National Natural Science Foundation of China [No. 51072159; 51273159], Science and technology program of Shaanxi Province [No: 2014K10-07], and Technology Foundation for Selected Overseas Chinese Scholar, Department of Human Resources and Social Security of Shaanxi Province [No: 2014-27]. The authors thank Jinyu Li and Tongmin Zhang (IMPCAS) for their help during the ion irradiation experiments.

Table S-1 Compositional analysis of different components of O 1s peak with increasing irradiation influence from XPS study

| Sample | Binding energy | | Area | | FWHM | |
|---|---|---|---|---|---|---|
| | O1s Scan I | O1s Scan II | O1s Scan I | O1s Scan II | O1s Scan I | O1s Scan II |
| S0 | 530.84 | 532.16 | 153265.61 | 25082.51 | 1.53 | 2.05 |
| S1 | 530.58 | 531.76 | 174981.72 | 60970.74 | 1.60 | 2.01 |
| S2 | 530.50 | 531.53 | 165186.14 | 79195.75 | 1.53 | 2.05 |
| S3 | 530.53 | 531.75 | 157277.63 | 56591.2 | 1.53 | 1.91 |
| S4 | 530.43 | 531.55 | 166159.04 | 72659.33 | 1.46 | 1.94 |
| S5 | 530.38 | 531.60 | 107606.75 | 68652.1 | 1.46 | 1.96 |